\documentclass[numberedappendix]{emulateapj}
\usepackage{epsfig}
\usepackage{graphicx}
\usepackage{color}
\usepackage{enumerate}
\newcommand{\kms}{\,{\rm km\,s^{-1}}}

\newcommand{\beq}{\begin{equation}}
\newcommand{\eeq}{\end{equation}}
\newcommand{\ba}{\begin{eqnarray}}
\newcommand{\ea}{\end{eqnarray}}
\def\spose#1{\hbox to 0pt{#1\hss}}
\newcommand{\lta}{\mathrel{\spose{\lower 3pt\hbox{$\mathchar"218$}}
      \raise 2.0pt\hbox{$\mathchar"13C$}}}
\newcommand{\gta}{\mathrel{\spose{\lower 3pt\hbox{$\mathchar"218$}}
      \raise 2.0pt\hbox{$\mathchar"13E$}}}
\def\msun{\,{\rm M_\odot}}
\newcommand\rsun{{\rm R_\odot}}

\def\ndot{\,{\rm yr^{-1}}}
\lefthead{Chen, Madau, Sesana, \& Liu}
\righthead{Tidal disruption of bound stars by MBH binaries}

\begin{document}
\submitted{}
\title{Enhanced tidal disruption rates from massive black hole binaries}
 
\author{Xian Chen\altaffilmark{1,2}, Piero Madau\altaffilmark{2}, Alberto Sesana\altaffilmark{3}, \& F. K. Liu\altaffilmark{1,4}}

\altaffiltext{1}{Department of Astronomy, Peking University, 100871 Beijing, China.}
\altaffiltext{2}{Department of Astronomy \& Astrophysics, University of California, Santa Cruz, CA 95064.}
\altaffiltext{3}{Center for Gravitational Wave Physics, The Pennsylvania State University, University 
Park, State College, PA 16802.}
\altaffiltext{4}{Kavli Institute for Astronomy and Astrophysics, Peking University, 100871 Beijing, China}

\begin{abstract}
``Hard" massive black hole (MBH) binaries embedded in steep stellar cusps can shrink
via three-body slingshot interactions. We show that this process will inevitably be accompanied 
by a burst of stellar tidal disruptions, at a rate that can be several orders of magnitude larger 
than that appropriate for a single MBH. Our numerical scattering experiments reveal that: 1) 
a significant fraction of stars initially bound to the primary hole are scattered into its tidal 
disruption loss cone by      gravitational interactions with the secondary hole, an enhancement effect 
that is more pronounced for very unequal-mass binaries; 2) about 25\% (40\%) of all strongly 
interacting stars are tidally disrupted by a MBH binary of mass ratio $q=1/81$ ($q=1/243$) 
and eccentricity 0.1; and 3) two mechanisms dominate the fueling of the tidal 
disruption loss cone, a Kozai non-resonant interaction that causes the secular evolution of the stellar 
angular momentum in the field of the binary, and the effect of close encounters with the secondary 
hole that change the stellar orbital parameters in a chaotic way. For a hard MBH binary of 
$10^7\,\msun$ and mass ratio $10^{-2}$, embedded in an isothermal stellar cusp of velocity dispersion
$\sigma_*=100\,\kms$, the tidal disruption rate can be as large as ${\dot N}_*\sim 1\,$ yr$^{-1}$. 
This is 4 orders of magnitude higher than estimated for a single MBH fed by two-body relaxation.
When applied to the case of a putative intermediate-mass black hole inspiraling onto 
Sgr A$^*$, our results predict tidal disruption rates ${\dot N}_*\sim 0.05-0.1\,$ yr$^{-1}$.
\end{abstract}


\section{Introduction}

Close massive black hole (MBH) binaries are expected to form in large numbers following 
the hierarchical assembly of massive galaxies \citep[e.g.][]{begelman80,volonteri03},
but their merger history remains poorly understood. Few observational probes of the processes 
that lead to and accompany the shrinking and inspiral of a MBH binary have been proposed to date: 1)
the gravitational slingshot ejection of hypervelocity stars from the Galactic Center into the halo
\citep[e.g.][]{yu03,levin06,sesana06,sesana07,brown09,perets09};  2) interruption or redirection of jets due to MBH binary-accretion disk interaction or MBH coalescence \citep{merritt02,liu03,liu04,liu07}; 
3) the coalescence of MBH pairs with masses in the range $(10^4-10^7)/(1+z)\,\msun$ giving origin 
to gravitational wave events that are one of the primary targets for the planned {\it Laser Interferometer 
Space Antenna} \citep[{\it LISA}; e.g.][]{haehnelt94,hughes02,wyithe03,sesana04,sesana05}; 4)
the electromagnetic afterglow from a circumbinary accretion disk that would follow such coalescence
\citep[e.g.][]{milos05,dotti06,lippai08,shields08};  and 5) the high-velocity 
recoil experienced by the plunging binary due to the asymmetric emission of gravitational waves 
\citep[e.g.][and references therein]{baker08}. A recoiling hole that retains the inner parts 
of its accretion disk may have fuel for a long-lasting luminous phase along its trajectory,
and shine as an off-center AGN \citep[e.g.][]{madau04,blecha08,volonteri08}.

In this {\it Letter} we return to the dynamical processes that determine the decay of MBH binaries
in a stellar background, prior to the gravitational wave regime, and put forward another possible 
observational signature of close binaries in the nuclei of galaxies. Using results from scattering 
experiments we show that gravitational slingshot interactions between an unequal-mass ``hard" binary 
and a bound stellar cusp will inevitably be accompanied by a burst of stellar tidal disruptions, 
at a rate that can be {\it several orders of magnitude larger} than that appropriate to a single 
MBH fed by two-body relaxation.  The duration of the phase of enhanced tidal disruption is of 
the order of $10^4-10^5\,$ yr.

\section{Scattering experiments}

Analytical techniques have been used by \citet{ivanov05} to study the enhanced stellar disruption rates 
induced by the secular non-resonant interaction with a non-evolving MBH binary. Here, we perform 
detailed numerical experiments of the close encounters between stars and the pair of MBHs,  
collisions that perturb stellar orbits in a chaotic way and scatter stars initially bound to 
the primary MBH into its tidal disruption loss cone. Consider a MBH binary of mass 
$M=M_1+M_2=M_1(1+q)$ ($M_2\ll M_1$) and semi-major axis $a$, orbiting in a background of stars 
of mass $m_*$, radius $R_*$, and velocity dispersion $\sigma_*$. 
When $a\lta a_h\equiv GM_2/4\sigma_*^2$, the ``hard" binary loses orbital energy by three-body 
slingshot interactions 
\citep{quinlan96,sesana06,sesana07}. For unequal-mass pairs, the radius of influence $r_{\rm inf}
\equiv G(M_1+M_2)/(2\sigma_*^2)$ 
is much larger than the hardening radius $a_h$, and almost all interacting (low angular-momentum)
stars are bound to $M_1$.\footnote{Note that, for extreme mass ratios $q\ll 1$,
the encounter is essentially a two-body scattering as the star and the secondary hole
move in the static potential of the primary.}\ In the case of an isothermal stellar distribution around
$M_1$, the total stellar mass within $a_h$ is equal to $M_2/2$.  

\begin{figure*}
\plotone{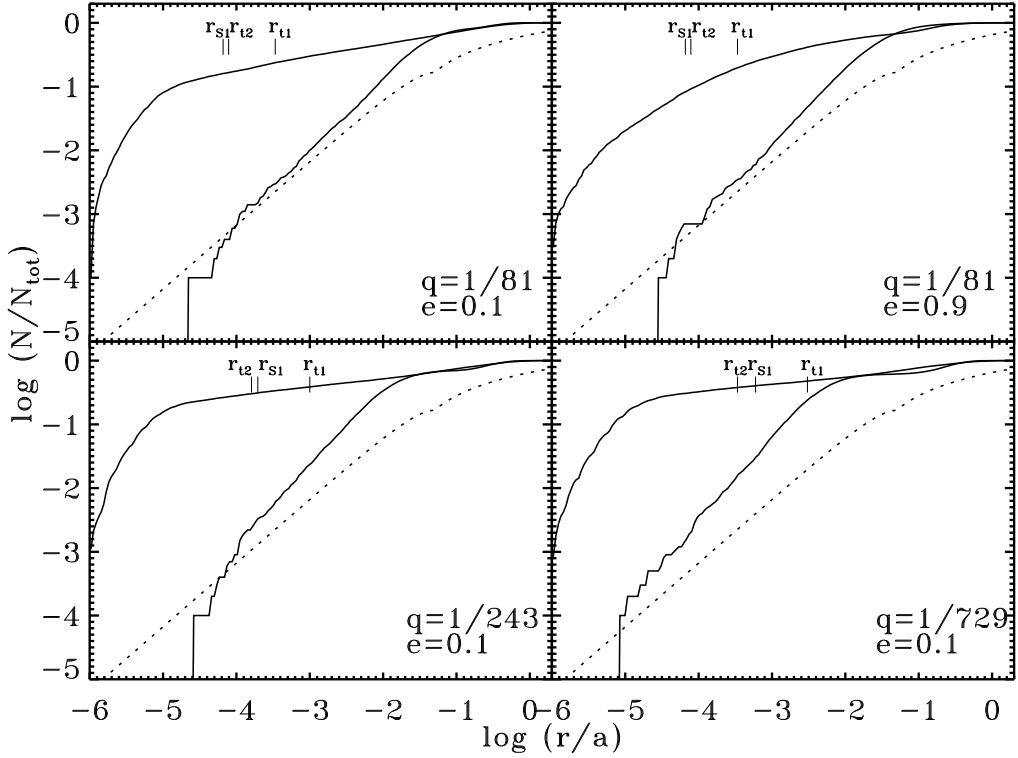}
\caption{{\it Upper solid line:} Close-encounter probability for bound stars 
interacting with the primary member (of mass $M_1$) of a MBH binary of mass ratio $q$, 
eccentricity $e$, and separation $a=a_h$. The vertical axis shows the fraction of 
stars $N/N_{\rm tot}$ with closest approach distance $r_{\rm min}<r$.      
{\it Lower solid line:} same for $M_2$. {\it Dotted line:} same in the case of an isolated MBH 
of mass $M_1$. Stars are drawn from a spherical isotropic distribution bound to $M_1$ and have semi-major 
axis in the range $a/2<a_*<2a$. The short vertical lines mark the tidal radii of $M_1$ and $M_2$ and the 
Schwarzschild radius of $M_1$ (all in units of $a$) for $M_1/m_*=10^7$, $r_*=\rsun$, and $\sigma_*=100\,\kms$.
From the top left to the bottom right, the fraction of stars tidally disrupted ($r_{\rm min}<r_{t1})$
by $M_1$ is 0.24, 0.19, 0.39, and 0.48, respectively.  
}
\label{fig1}
\vspace{0.2cm}
\end{figure*}

The integration of the three-body encounter equations is performed in a coordinate system 
centered at the location of $M_1$. Initially the binary (of mass ratio $q$) has eccentricity $e$ and 
a randomly-oriented orbit with $M_2$ at its pericenter. Stars initially move in the $x-y$ plane 
with pericenters along the positive $x$-axis and random orbital phases. The initial conditions 
of the restricted three-body problem problem are then completely defined by 6 variables, 3 for the
binary and 3 for the star: 1) the inclination of the binary orbit, $\theta$; 2) the longitude of 
$M_2$'s ascending node, $l$; 3) the argument of $M_2$'s pericenter, $\phi$; 4) the semi-major axis of 
the stellar orbit, $a_*$; 5) the specific angular momentum of the star, $j_*$; and 6) the orbital phase of 
the star, $p_*$. We start each scattering experiment by generating $6$ random numbers, with 
$\cos\theta$ evenly sampled in the range $[-1,1]$, and $l$ and $\phi$ uniformly distributed in the range 
$[0,2\pi]$. We sample $a_*$ logarithmically around $a$ in the range $[1/2a,2a]$ where three-body 
interactions are strongest, $j_*^2$ randomly between 0 and 1 
(corresponding to an isotropic distribution), and $p_*$ evenly between $0$ and $1$.
The equation of motion are integrated using an explicit Runge-Kutta method of order 8 \citep{hairer87}, 
with a fractional error per step in position and velocity 
set to $10^{-13}$. We have tested our code by reproducing Figures 4 and 6 in \citet{sesana08} and found
excellent agreement, and run $10^4$ scattering experiments for each binary configuration. During 
each experiment 
the minimum separation $r_{\rm min}$ between the star and the MBH pair is measured and stored: 
stars on orbits intersecting the tidal radius of hole $M_i$ ($i=1,2$),
\begin{equation}\label{rt}
r_{ti}=r_*\left(\frac{M_i}{m_*}\right)^{1/3}\\ \simeq (2.3\times10^{-6}~{\rm pc})
~\left(\frac{r_*}{\rsun}\right) \left(\frac{M_i}{10^6 m_*}\right)^{1/3},
\end{equation}
will be tidally disrupted at pericenter passage (neglecting general relativistic effects that
set in when $M_1\gg 10^7\,\msun$, see \citealt{hills75}). The results 
of our numerical experiments are shown in Figure \ref{fig1} for binaries with different mass 
ratios and eccentricities, all at separation $a=a_h$. The fraction of interacting stars 
that are scattered by $M_2$ to within a pericenter distance $r_{\rm min}<r_{t1}$ from $M_1$ and are tidal
disrupted can be, for very unequal mass binaries, orders of magnitude higher than the corresponding number
were $M_1$ not in a binary. The latter is simply given by all the bound stars within $M_1$ 
``tidal loss cone'', the region in the $(a_*-j_*)$ phase-space bounded by 
\begin{eqnarray}
j_{\rm lc}^2&=&
\left\{
\begin{array}{ll}
1\,\,\,\,\,\,\,\,(a_*<r_{t1})\\
2(r_{t1}/a_*)^2(a_*/r_{t1}-1/2)\,\,\,\,\,\,\,\,(a_*\ge r_{t1}),
\end{array}
\right.
\end{eqnarray}
where $j_*$ is the specific angular momentum of the star normalized to the angular momentum 
of a circular orbit with the same semi-major axis. For a binary with $q=1/81$, $e=0.1$, the probability 
that a close encounter with a star having $1/2a<a_*<2a$ results in a tidal disruption by $M_1$ 
($r_{\rm min}<3\times 10^{-4}a\approx r_{t1}$) is more than 2 orders of magnitude larger than if $M_1$ 
were single. The figure also shows that: 1) many more stars are disrupted by $M_1$ than by $M_2$; 2) the 
disruption probability decreases with increasing $q$. This is both because the ratio $r_{t1}/a_h$ decreases with 
increasing $q$, and because, as the perturbing force of the 
secondary increases, more stars are ejected altogether rather than disrupted; and 3) 
more stars are scattered into $M_1$'s tidal loss cone with decreasing binary eccentricity.

\begin{figure*}
\plotone{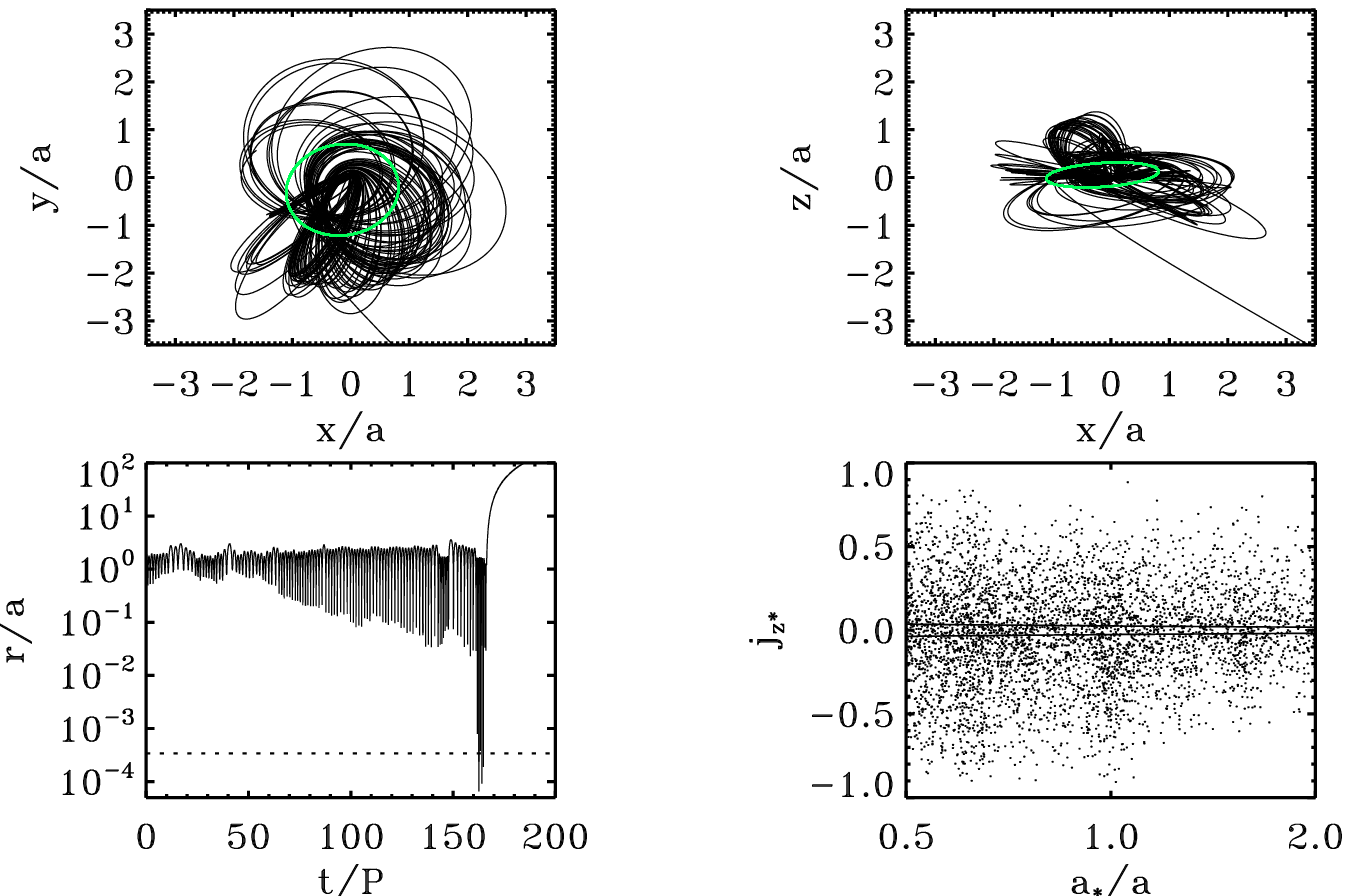}
\caption{{\it Top and botton left panels:} Example of a chaotic three-body scattering leading to a 
tidal disruption. A star on a bound orbit of semi-major axis $a_*=1.2a$ and 
eccentricity $e_*=0.5$ interacts with a MBH binary of parameters $M_1=10^7~\msun$, $q=1/81$, 
$e=0.3$, $a=a_h$. {\it Top left:} stellar ({\it black}) and $M_2$ ({\it green}) 
trajectories in the $x-y$ plane. The primary hole $M_1$ is located at the origin. {\it Top right:} same 
projected onto the $x-z$ plane. {\it Bottom left:} separation between the star and $M_1$ as a function of 
time (in units of the binary period $P$). The dotted line marks the tidal radius $r_{t1}$. {\it Bottom 
right panel:} Bound stars in the $(a_*-j_{z*})$ plane that are tidally disrupted during the interaction with a 
MBH binary of parameters $M_1=10^7~\msun$, $q=1/81$, $e=0.1$, $a=a_h$. The solid lines mark the
boundaries of the Kozai wedge $|j_{z*}|<j_{\rm lc}$.}
\label{fig2}
\vspace{0.8cm}
\end{figure*}

\section{Basic theory and disruption rates}

Figure \ref{fig2} shows an example of a three-body scattering leading to a tidal disruption 
after many pericenter passages. In the presence of a secondary black hole, strongly interacting
($a_*\sim a$) stars in nearly circular, highly inclined orbits relative to the binary orbital 
plane undergo 
a slow secular evolution that periodically increases their eccentricity (in exchange for a 
lower inclination) and eventually drives them within the tidal disruption loss cone of 
the primary hole \citep{ivanov05}. Our scattering experiments
show that this mechanism -- analogous to the so-called ``Kozai effect" of celestial 
mechanics \citep{kozai62} -- contributes but does not dominate the fueling of the tidal disruption
loss cone in the case of very unequal mass binaries. The stars supplied to the disruption loss 
cone by the Kozai effect are all those having normal 
component of the angular momentum, $j_{z*}$, within the loss cone, i.e. all those stars within a 
wedge-like region in phase space where $|j_{z*}|<j_{\rm lc}$. For a given stellar semi-major axis, 
$a_*\ll a$, the fraction of stars that lie outside the tidal disruption loss cone but inside the 
``Kozai wedge'' is 
\begin{equation}
f_K(r_{t1},a_*)=\int_{j_{\rm lc}}^{1}dj_*\int_{-j_{\rm lc}}^{j_{\rm lc}}dj_{z*}=
2j_{\rm lc}-2j_{lc}^2.
\end{equation}
When $r_{t1}\ll a_*$, $f_K\simeq 2\sqrt{2r_{t1}/a_*}$, which is $\sqrt{2a_*/r_{t1}}$ times larger 
than the fraction of stars already in the tidal loss cone, $j_{lc}^2\simeq 2r_{t1}/a_*$. This result 
explains why the probability of stellar disruption for bound stars is much higher in MBH binaries 
than in single black hole systems. It also predicts that only a fraction $2\sqrt{2r_{t1}/a_*}=
(0.05,0.09,,0.15)$ 
for $q=(1/81,2/243,1/721)$ of all stars with $a/2<a_*<2a$ will be supplied to the tidal loss cone 
by the Kozai effect. Figure \ref{fig1} shows that for very unequal mass binaries the tidal disruption 
probability is much larger than the above estimate. This discrepancy highlights the importance of 
close, resonant encounters with the secondary hole, which change the stellar orbital parameters in a 
chaotic way and fuel the tidal loss cone. Figure \ref{fig2} (bottom right panel) depicts the initial 
distribution in the $(a_*-j_{z*})$ plane of all the stars that are disrupted in our numerical 
experiments. It is clear that the majority of disrupted stars lie outside the Kozai wedge. 

\begin{figure*}
\plotone{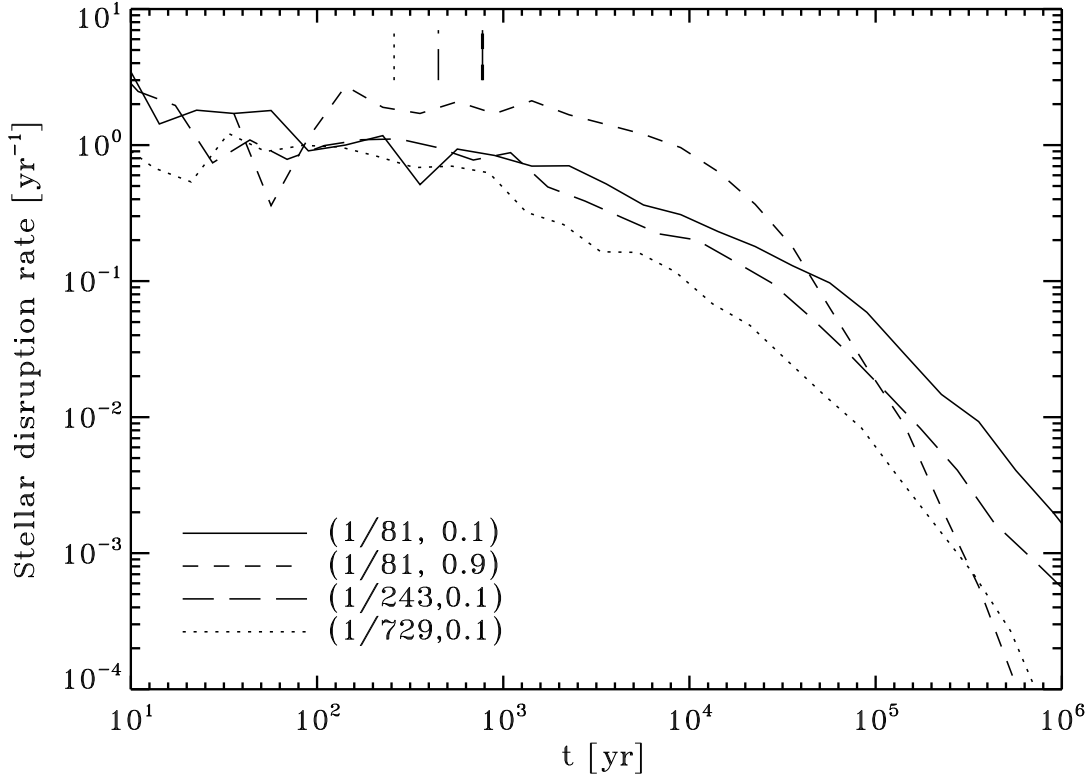}
\caption{Numerical tidal disruption rates as a function of time for a hard MBH binary 
of mass ratio $q$, eccentricity $e$, and separation $a=a_h$, embedded in an isothermal stellar cusp. 
The derivation assumes $M_1=10^7\,\msun$, $m_*=1\,\msun$, $r_*=\rsun$, and $\sigma_*=100\,\kms$.
The short vertical lines mark the Kozai timescale for $a_*=a=a_h$.
}
\label{fig3}
\end{figure*}

We now show that the ejection of ambient stars by a hard MBH binary
will be accompanied by a burst of stellar tidal disruptions, at a rate that may be 
orders of magnitude larger than that appropriate for a single MBH. In our numerical 
experiments the time when a star first crosses the tidal radius of the primary hole 
is stored and used to calculate a disruption frequency. To translate this number into a stellar 
disruption rate in physical units one needs to specify the parameters of the MBH binary and its 
stellar cusp. At the hardening radius 
\begin{equation}
a_h\equiv \frac{GM_2}{4\sigma_*^2}\simeq (1.1{~\rm pc})~\sigma_{100}^{-2}q~M_7
\end{equation}
the orbital period of the binary is
\begin{equation}\label{pah}
P_h=2\pi\sqrt{\frac{a_h^3}{G(M_1+M_2)}}\simeq (3.3\times10^4~{\rm yr})~
\sigma_{100}^{-3} M_7\left (\frac{q^3}{1+q}\right)^{1/2}, 
\end{equation}
where $\sigma_{100}\equiv \sigma_*/100\,\kms$ and $M_7\equiv M_1/10^7\,\msun$. 
If we assume now, for simplicity, that the stars bound to $M_1$ follow an isothermal distribution, 
$\rho_*(r)=\sigma_*^2/(2\pi Gr^2)$, the total stellar mass between $a_h/2$ and $2a_h$ is $M_*=3qM_1/4$.
Normalizing the interacting mass in our numerical experiment to the isothermal case and rescaling 
the stored disruption times according to equation (\ref{pah}), we obtain the stellar disruption rates shown in Figure~\ref{fig3}.   

Although the standard Kozai theory does not strictly apply to strongly interacting stars, we use it here to 
derive an analytical scaling for the stellar disruption rates. The Kozai timescale at $a_h$ is approximately
\begin{equation}\label{tk}
T_{\rm K}=\frac{2}{3\pi q}\left(\frac{a_*}{a_h}\right)^{-3/2}P_h
\end{equation}
\citep{innanen97,kiseleva98}. The stellar disruption rate can then be estimated as
\begin{equation} 
{\dot N}_*=\frac{\lambda f_K M_*}{m_*T_K}e^{-t/T_K} \nonumber
\end{equation}
\begin{equation}
~~~~~~~~\simeq (6~\ndot)~\lambda(1+q)^{1/2}\sigma_{100}^4M_7^{-1/3}e^{-t/T_K},
\label{drate}
\end{equation}
where $f_K$ is the fraction of stars in the Kozai wedge and $\lambda\simeq0.2$ is a correction factor accounting for the 
uncertainty in $T_K$ and for stars that are actually ejected before disruption, as well as for properly weighting our scattering 
experiments for the case of an isothermal profile. The numbers provided
by equation (\ref{drate}) are in good agreement with the numerical rates.
The tidal disruption plateau, however, lasts much longer than $T_K$ (indicated by the vertical ticks 
in Fig. \ref{fig3}), because at late times the disruption rate is dominated by chaotic scatterings. 

The tidal disruption rates we compute are many orders of magnitude higher than those, 
\beq
\dot{N}_*\simeq (2\times10^{-4}~\ndot)~\sigma_{100}^{7/2}M_7^{-1}\left(\frac{m_*}{\msun}\right)^{-1/3}
\left(\frac{r_*}{\rsun}\right)^{1/4}, 
\eeq
derived for a single MBH fed by two-body relaxation \citep{wang04}. Note that in our calculations 
we have only considered stars with $a/2<a_*<2a$. Taking into account stars in a larger range of 
semi-major axis would further boost the binary disruption rates.

\section{Conclusions}

We have used results from numerical scattering experiments and shown that the tidal disruption rate
in a stellar cusp containing a $10^7\,\msun$ MBH binary can be as large as $1\,\ndot$ over a timescale
of $\sim 10^5\,$yr. {\it This is orders of magnitude larger than expected in the case of single MBHs}.  
After a tidal disruption, about half of the debris will be spewed into eccentric bound orbits and 
fall back onto the hole, giving rise to a bright UV/X-ray outburst that may last for a few years 
\citep[e.g.][]{Rees1988}. ``Tidal flares" from MBHs may have been observed in several nearby 
inactive galaxies 
\citep{komossa02,esquej07}. The inferred stellar disruption frequency is $\sim10^{-5}~{\rm yr^{-1}}$ 
per galaxy (with an order of magnitude uncertainty, \citealt{donley02}).
The much enhanced disruption rates we have found here for MBH binaries can 
then be used to constrain the abundance of close MBH pairs in nearby galaxy nuclei \citep{chen08}. 
It is interesting to scale our results to the scattering of stars bound to Sgr A$^*$,
the massive black hole in the Galactic Center, by a hypothetical inspiraling companion of intermediate
mass \citep{yu03,sesana07}. The stellar density profile around the Galactic Center can be described
as a double power-law, with outer slope $\simeq-2$ and inner slope $\simeq-1.5$ \citep{schodel07}. 
If the density profile inside the influence radius of $M_1$ is shallower than isothermal, 
\beq
\rho_*(r<r_{\rm inf})=\rho_*(r_{\rm inf})\left(\frac{r}{r_{\rm inf}}\right)^{-\gamma}
\eeq
with $\gamma<2$, then the stellar mass between $2a_h$ and $a_h/2$ decreases by a 
factor $(a_h/r_{\rm inf})^{2-\gamma}$, and the stellar disruption rate in equation (\ref{drate})
decreases by a factor $(q/4)^{2-\gamma}$ relative to the isothermal case.
Using $M_1=4\times 10^6\,\msun$, $\sigma_*=100\,\kms$, $\gamma=1.5$, and $1/243<q<1/81$, yields 
rates in the range ${\dot N}_*\simeq 0.05-0.1\,{\rm yr^{-1}}$. 

There are a number of uncertainties in our calculations that require clarification 
before a firm statement can be made on the rates and duration of stellar tidal disruptions 
expected in galaxy nuclei hosting MBH binaries, and on the constraints imposed by the 
very low level of activity observed in the Galactic Center. First and foremost,  
our estimates of the tidal disruption rate assume a fixed binary separation $a$. 
In reality, both binary separations and eccentricities will evolve due to three-body slingshots 
\citep{sesana08}. This changes the Kozai timescale of the system and replenishes the suppply of 
strongly interacting stars. According to Figure~7 of \citet{sesana08}, the evolutionary timescale 
of a binary with initial eccentricity 0.1 embedded in an isothermal cusp is $t_h\sim q^{-3/2}P$.
At $a=a_h$ we derive
\beq\label{th}
t_h\sim (3.3\times10^4~{\rm yr})~\sigma_{100}^{-3} M_7(1+q)^{-1/2}.
\eeq
This is comparable to the duration of the plateau in the disruption rates shown in 
Figure \ref{fig3}, implying that binary evolution should not qualitatively change 
the plateau values. A more
sophisticated calculation that couples the results of numerical scattering experiments 
with an evolving binary will be the subject of a subsequent paper. 

\acknowledgments
Support for this work was provided by NASA through grant NNX08AV68G (P.M.). X.C. and F.K.L. thank the Chinese national 973 program (2007CB815405) 
and the China Scholarship Council for financial support. We are grateful to F. Haardt for early discussions on this topic.

\end{document}